\documentclass[english]{article}
\usepackage[T1]{fontenc}
\usepackage[latin1]{inputenc}

\makeatletter

\usepackage{babel}
\makeatother
\begin{document}

\title{Modified Grover's search algorithm for the cases where the number
of solutions is known }

\maketitle
\begin{center}Abhishek S Gupta%
\footnote{acmabhi@rediffmail.com%
}, Manu Gupta%
\footnote{manu\_friends@yahoo.com%
}, Anirban Pathak%
\footnote{anirban.pathak@jiit.ac.in%
} \end{center}

\begin{center}Jaypee Institute of Information Technology, A-10, Sector
62, Noida-201307, India \end{center}

\begin{abstract}
Grover's search algorithm searches a database of $N$ unsorted items
in $O(\sqrt{N/M})$ steps where $M$ represents the number of solutions
to the search problem. This paper proposes a scheme for searching
a database of $N$ unsorted items in $O(logN)$ steps, provided the
value of $M$ is known. It is also shown that when $M$ is unknown
but if we can estimate an upper bound of possible values of $M$,
then an improvement in the time complexity of conventional Grover's
algorithm is possible. In that case, the present scheme reduces the
time complexity to $O(MlogN)$. 
\end{abstract}

\section{Introduction}

With the advent of quantum computation many quantum algorithms {[}\ref{the:M.A.-Nielsen-and}-\ref{the:L.-K.-Grover,}{]}
which, work faster than their classical counter parts, have appeared.
Among these quantum algorithms, Grover's algorithm {[}\ref{the:L.-K.-Grover,}{]}
got special attention of the whole community because of its wide applicability
in searching databases. Actually, searching databases is one of the
most important problems in computer science and real life. This fact
has motivated people to develop a large number of algorithms to search
different kind of databases {[}\ref{the:coreman}{]}. We are interested
about a database of $N$ unsorted items, having $M$ solutions (where
$M\ll N)$. Any classical algorithm takes $O(N)$ steps to search
such a database. Grover's quantum algorithm {[}\ref{the:L.-K.-Grover,}{]}
searches such a database in $O(\sqrt{N/M})$ steps. Till now, the
time complexity of Grover's algorithm is minimum among all the algorithms
designed for the same purpose. Tight bound on Grover searching has
been studied by many people {[}\ref{tight bound}{]} but it does not
establish any tight bound on quantum search in general. This fact
has motivated us to explore the possibility of improvement in special
cases. In Grover's original algorithm the number of solution $M$
is unknown. In the present work we consider a special case of Grover's
algorithm and assume that either $M$ is known or an upper bound of
$M$ is known. In the first case time complexity reduces to $O(logN)$
and in the second case it reduces to $O(MlogN)$. The reduction is
considerably large when $M$ is small and $N$ is large. 

In section 2 we briefly discuss Grover's algorithm. The modified algorithm,
is discussed in section 3. In section 4 we have discussed time complexity
of various cases. Finally we conclude in section 5.

\section{Grover's algorithm}

As we have already stated, we are interested to search a database
of $N$ items out of which $M$ are the solutions. In Grover's search
algorithm we assign an index to each element and search on those indices.
Now for our convenience if we assume that $N=2^{n}$ then we can store
all the indices in $n$ qubits since the indices varies from $0$
to $N-1$. A particular instance of the search problem can conveniently
be represented by a function $f$, which takes an integer $x$, in
the range $0$ to $N-1$. By definition, $f(x)=1$ if $x$ is a solution
to the search problem and $f(x)=0$ if $x$ is not a solution to the
search problem. 

Grover's algorithm uses an unitary operator as a quantum oracle which
flips the oracle qubit if $f(x)=1$. Essentially, the Oracle marks
the solutions to the search problem by shifting the phase of the solution.
The search oracle is applied only $O(\sqrt{N/M})$ times in order
to obtain a solution on a quantum computer. This is done in following
steps:

\begin{enumerate}
\item The algorithm begins with a quantum register in the state $\left|0\right\rangle ^{\otimes n}$.
\item The Hadamard transform is used to put the register in a equal superposition
of $N=2^{n}$ states. This is how we used to prepare the input state
$|x>$ for the oracle.
\item A quantum subroutine, known as the Grover iteration is repeatedly
applied. The Grover iteration may be broken in following four steps:\\
a) Apply the oracle \\
b) Apply the Hadamard transformation on $n$ qubits \\
c) Perform a conditional phase shift on the computer, with every computational
basis state except $|0\rangle$ receiving a phase shift of $-1$\\
d) Apply the Hadamard transformation on $n$ qubits. 
\end{enumerate}

\section{The modified algorithm }

To simplify the understanding of Grover's algorithm, we can assume
that the initial superposition is constituted of 2 parts: the solution
states and the non-solution states and represent the state as \begin{equation}
|S\rangle=\cos\theta|0\rangle>+\sin\theta|1\rangle\label{eq:s}\end{equation}
where $|0\rangle$ represents the non-solution states and $|1\rangle$
represents the solution states and\[
\cos\theta=\sqrt{\frac{(N-M)}{N}}\]

\[
\sin\theta=\sqrt{\frac{M}{N}}.\]
The Oracle can be considered as a $2\times2$ matrix which flips the
phase of the solution states. It can be written as : \[
O=\left(\begin{array}{cc}
1 & 0\\
0 & -1\end{array}\right).\]

At the end of each Grover iteration (Grover iteration is a phase flip
of the solution states, followed by an inversion of all states about
the mean), the initial state gets rotated by an angle of $2\theta$
in a direction such that it moves closer to the solution states. In
other words, each Grover iteration increases the probability of the
solution states (simultaneously decreasing the probability of the
non-solution states). Therefore, the correct solution can be measured
with a high probability after a certain number of Grover iterations.
Essentially, a particular Grover iterator redistributes the probability
among the possible states in two steps. First it flips the phase of
the solution states and then inverts about the mean. In this process,
the probability of nonsolution states gets reduced and the reduced
probability is added to those of the solution states. Here an important
question arises: Is it essential to invert the states about mean?
The answer is no! Actually, the essential condition is unitarity of
the operation. When $M$ is unknown, then this one of the unitary
operation through which we can redistribute probability according
to the requirement and conserve the total probability. So when $M$
is unknown the state represented by (\ref{eq:s}) has to be rotated
by $2\theta$ in each step. But if we know the, value of $M$ (i.e.
we know $\theta$), then we can introduce an unitary operation which
vanishes the probability of appearance of nonsolution states and uniformly
distributes that probability among all the solution states. This new
unitary operation exploits the fact that if $M$ is known then the
amount of rotation which can map the initial state into the solution
state is known. Geometrically, these means an inversion about a suitable
point (instead of the inversion about the mean).

The equation to determine the number of iterations $I$ required in
conventional Grover's algorithm is \begin{equation}
\theta+I(2\theta)=\frac{\pi}{2}.\label{eq:suru}\end{equation}
Now, instead of carrying out Grover's iteration large number of times,
we propose carrying out the same action in one step i.e. instead of
rotating the current search state by $2\theta$, we propose rotating
it directly by $k\theta$ where: \begin{equation}
\theta+k\theta=\frac{\pi}{2}\label{eq:mod1}\end{equation}
i.e. \begin{equation}
k\theta=\frac{\pi}{2}-\theta.\label{eq:mod2}\end{equation}
 Thus, if we can rotate the current search state (initial state) by
$k\theta$ then we can obtain the desired solution state in a single
iteration. The time complexity of the process is $O(logN)$ (to create
Hadamard superposition). 

A $2\times2$ matrix that rotates a state vector (represented by a
$2\times1$ matrix in 2 dimensions) by $k\theta$ can be written as:
\[
\left(\begin{array}{cc}
cosk\theta & -sink\theta\\
sink\theta & cosk\theta\end{array}\right).\]
Replacing $k\theta$ by $\frac{\pi}{2}-\theta$ we get a new operator
$A$ defined as follows: 

\begin{equation}
A=\left(\begin{array}{cc}
sin\theta & -cos\theta\\
cos\theta & sin\theta\end{array}\right).\label{eq:mod3}\end{equation}
To understand the physical meaning of this operation let us assume
that the rotation operation A is obtained as the oracle operation
O followed by another operation (say X), i.e. XO=A. Solving the above
equation, we get X as: \begin{equation}
X=\left(\begin{array}{cc}
sin\theta & cos\theta\\
cos\theta & -sin\theta\end{array}\right).\label{eq:mod4}\end{equation}
The matrix X can be written in operator form as: \begin{equation}
X:=\left(sin\theta|0\rangle+cos\theta|1\rangle\right)\langle0|+\left(cos\theta|0\rangle-sin\theta|1\rangle\right)\langle1|.\label{eq:mod5}\end{equation}

According to our basic assumption the value of $M$ and $N$ are known.
Therefore, $cos\theta$ and $sin\theta$ are known and we can prepare
the unitary operation $X$. It is easy to check that $X$ is unitary
and physically $X:$ represents a quantum gate which causes an inversion
about a point such that the nonsolution state probabilities are reduced
to zero. The $X$ can be multiplied with $O$ to produce $A$, which
operates on $|S\rangle$as follows, \begin{equation}
A:|S\rangle=A:\left(\begin{array}{c}
cos\theta\\
sin\theta\end{array}\right)=\left(\begin{array}{c}
0\\
1\end{array}\right).\label{eq:mod7}\end{equation}
 Thus, we are left only with solution states that can be obtained
by performing a measurement. Essentially, the modification of the
point of inversion reduces the time complexity in our case. But only
if we know the total number of solutions then we can choose the suitable
point of inversion.

\section{Time complexity in various cases: }

\begin{enumerate}
\item M (the number of solutions to the search query) is known\\
Only 1 iteration is required to reach the solution state. The input
is prepared by applying appropriate number of Hadamard gates resulting
in equal superposition of N states. Thus, the total time complexity
would be $O(logN)$.
\item M is unknown but we can estimate an upper bound on the possible value
of M. \\
The algorithm suggested above can only be executed only for a particular
value of M. Since we are aware of the upper bound, measurement of
the register which holds the answer (for a particular value of M)
is checked to be correct. Thus, for each value of M we are required
to verify the correctness of the answer provided by running the algorithm.
The answers obtained can be checked easily as stated in {[}\ref{tight bound}{]}.
This approach will lead to a time complexity of $O(MlogN)$.
\item M is unknown and we cannot estimate an upper bound on the possible
value of M. \\
An alternative approach can be adopted in this case. Recently we have
given a proposal {[}\ref{the:gupta-pathak1}{]} to handle this case
using concurrency control techniques and marking. This proposal reduces
the complexity to $O(M+logN)$.
\end{enumerate}

\section{Conclusion }

This paper proposes a scheme to search a database of $N$ unordered
items in $O(logN)$ when $M$ is known. And in $O(MlogN)$ when $M$
is unknown but an estimation of upper limit of $M$ is possible. This
improvement in complexity is considerable and it will be more prominent
with the increase of the size of the database. There exist many applications
of Grover's algorithm. Thus, an improvement in Grover's search will
result in the improvement in time complexity of all these applications.
This is a special case of Grover's algorithm where time complexity
is less than that of the conventional Grover's algorithm. There may
exist many similar special cases of more general quantum search problem
where complexity is less. Thus the present study opens up a possibility
to look at the quantum search problems from a new perspective.


\begin{thebibliography}{1}
\bibitem{key-1}\label{the:M.A.-Nielsen-and}M. A. Nielsen and I. L. Chuang, \emph{Quantum
Computation and Quantum Information}, Cambridge University Press (2000).
\bibitem{key-5}\label{the:p.shor} P. W Shor, \emph{Polynomial time algorithm for
Prime Factorisation and Discrete Logarithms on a Quantum Computer},
quant-ph/9508027 (1994).
\bibitem{key-8}\label{the:Gupta Pathak old one}A. S. Gupta and A. Pathak, \emph{Quantum
Flyod Warshall},quant-ph/0502144 (2005).
\bibitem{key-2}\label{the:L.-K.-Grover,}L. K. Grover, \emph{A fast quantum mechanical
algorithm for database Search}, Proceedings, 28th Annual ACM Symposium
on the Theory of Computing (STOC), 212 (1996).
\bibitem{key-6}\label{the:coreman}T. H. Cormen, C. E. Leiserson, R. L. Rivest, C.
Stein, \emph{Introduction to Algorithms}, MIT Cambridge (2001).
\bibitem{key-3}\label{tight bound}M. Boyer, G. Brassard, P. Hoyer, A. Tapp, \emph{Tight
Bounds on Quantum Searching}, Fortschr. Physics \textbf{46} 493 (1998).
\bibitem{key-4}\label{the:gupta-pathak1}A. S. Gupta and A. Pathak, \emph{Modified
Grover's search algorithm in$O(M+logN)$}, quant-ph/0506093 (2005).\end{thebibliography}
\end{document}